\documentclass[journal,a4paper]{IEEEtran}
\pdfoutput=1

\usepackage[T1]{fontenc}
\usepackage[utf8]{inputenc}

\usepackage{amsmath,amssymb}
\usepackage{booktabs}
\usepackage{comment}
\usepackage{cite}
\usepackage{enumitem}
\usepackage{graphicx}
\usepackage{lipsum}
\usepackage{xcolor}
\usepackage{xurl}

\usepackage[acronym]{glossaries}
\glsdisablehyper

\newacronym{afxdp}{AF\_XDP}{Address Family eXpress Data Path}
\newacronym{api}{API}{application programming interface}
\newacronym{asic}{ASIC}{application-specific integrated circuit}
\newacronym{axioe}{AXIoE}{AXI over Ethernet}
\newacronym{bth}{BTH}{base transport header}
\newacronym{chdr}{CHDR}{Condensed Hierarchical Datagram for RFNoC}
\newacronym{cm}{CM}{connection manager}
\newacronym{cpu}{CPU}{central processing unit}
\newacronym{difi}{DIFI}{Digital Intermediate Frequency Interoperability}
\newacronym{dma}{DMA}{direct memory access}
\newacronym{dpdk}{DPDK}{Data Plane Development Kit}
\newacronym{dpu}{DPU}{data processing unit}
\newacronym{dram}{DRAM}{dynamic random-access memory}
\newacronym{ecpri}{eCPRI}{enhanced Common Public Radio Interface}
\newacronym{felix}{FELIX}{Front-End Link eXchange}
\newacronym{fpga}{FPGA}{field-programmable gate array}
\newacronym{fspl}{FSPL}{free-space path loss}
\newacronym{gbe}{GbE}{Gigabit Ethernet}
\newacronym{icrc}{ICRC}{invariant CRC}
\newacronym{ipv4}{IPv4}{Internet Protocol version 4}
\newacronym{iq}{I/Q}{in-phase and quadrature}
\newacronym{isac}{ISAC}{integrated sensing and communication}
\newacronym{iwarp}{iWARP}{Internet Wide Area RDMA Protocol}
\newacronym{mtu}{MTU}{maximum transmission unit}
\newacronym{nic}{NIC}{network interface card}
\newacronym{numa}{NUMA}{non-uniform memory access}
\newacronym{pcie}{PCIe}{Peripheral Component Interconnect Express}
\newacronym{psn}{PSN}{packet sequence number}
\newacronym{qpn}{QPN}{queue pair number}
\newacronym{qp}{QP}{queue pair}
\newacronym{rcs}{RCS}{radar cross section}
\newacronym{rdma}{RDMA}{remote direct memory access}
\newacronym{reth}{RETH}{RDMA extended transport header}
\newacronym{rfnoc}{RFNoC}{RF Network-on-Chip}
\newacronym{rocev2}{RoCEv2}{RDMA over Converged Ethernet version 2}
\newacronym{rx}{Rx}{receiver}
\newacronym{sdr}{SDR}{software-defined radio}
\newacronym{shampo}{SHAMPO}{split header and merge payload offload}
\newacronym{snr}{SNR}{signal-to-noise ratio}
\newacronym{tx}{Tx}{transmitter}
\newacronym{udp}{UDP}{User Datagram Protocol}
\newacronym{uhd}{UHD}{USRP Hardware Driver}
\newacronym{usb}{USB}{Universal Serial Bus}
\newacronym{vita49}{VITA 49}{VITA Radio Transport}

\usepackage{hyperref}
\hypersetup{
	pdfauthor={Carsten Andrich, Sebastian Giehl, Michael Schubert},
	pdftitle={Beyond UDP: RDMA as the Missing Link for Gigasample Software-Defined Radio},
	citebordercolor={.8 .8 .8},
	linkbordercolor={.8 .8 .8},
	urlbordercolor={.8 .8 .8},
	pdfborderstyle={/S/U/W 1}
}

\usepackage{orcidlink}

\usepackage{tikz}
\usetikzlibrary{positioning,calc}
\usetikzlibrary{arrows.meta,shapes}

\title{Beyond UDP: RDMA as the Missing Link\\for Gigasample Software-Defined Radio}

\author{%
    \IEEEauthorblockN{%
        Carsten~Andrich\raisebox{.5ex}{\orcidlink{0000-0002-4795-3517}},
        Sebastian~Giehl\raisebox{.5ex}{\orcidlink{0009-0008-1672-1351}},
        Michael~Schubert\raisebox{.5ex}{\orcidlink{0009-0003-4843-2507}}%
    }%
    \thanks{
        The research was funded by the German Federal Ministry for Economic Affairs and Energy and the European Union through the European Social Fund Plus (ESF Plus) under grant 03EFTH0036 (project ``unparallabs'').
        The research was also funded by the German Federal Ministry of Research, Technology and Space under grant 16KIS2225 (project ``6Gsens'').

        All authors are with the Institute of Information Technology, Technische Universität Ilmenau, Ilmenau, Germany (e-mail: \href{mailto:carsten.andrich@tu-ilmenau.de}{carsten.andrich@tu-ilmenau.de}).
        The authors intend future commercial exploitation of the described method.

    }
}

\begin{document}

\maketitle

\begin{abstract}
Software-defined radio (SDR) front ends have gained an order of magnitude in sample rate within a single product generation, while the transport that carries their output has not changed: samples are packed into Ethernet datagrams and reassembled by the host processor.
At gigasample rates this fails on two counts at once, exhausting the time the host has per packet and the memory bandwidth it has for copying, and no amount of tuning or kernel bypass recovers both.
The burden has instead been passed to the user, who must supply operating system expertise that has nothing to do with radio.
We show that remote direct memory access removes the problem rather than relocating it.
The SDR places received samples directly into memory the host has registered, and fetches samples for transmission from that memory when its converters require them.
The processor leaves the data path in both directions, each sample byte crosses the memory bus once, and the transmit path loses the hard real-time pacing problem that has always accompanied it.
What remains for the user application is the signal processing itself, which should be all an SDR requires, even at gigasample rates.
\end{abstract}

\begin{IEEEkeywords}
Analog-digital conversion, digital-analog conversion, Ethernet networks, field programmable gate arrays, InfiniBand, memory management, network interfaces, real-time systems, remote direct memory access (RDMA), RDMA over Converged Ethernet (RoCE), software radio.
\end{IEEEkeywords}

\section{Introduction}
\label{sec:intro}

\IEEEPARstart{S}{ample} rates of \gls{sdr} front ends have risen by an order of magnitude within a single product generation.
A USRP X310 delivers 200\,MSa/s per channel, or 0.8\,GB/s.
A USRP X440 delivers 2\,GSa/s, or 8\,GB/s.
This technological leap was facilitated by the Xilinx RFSoC platform, and even more capable hardware is already in early access.
The Xilinx Versal RF series supports up to 16\,GSa/s, or 64\,GB/s from a single channel and comparable Altera products field similar numbers.
The transport that carries these samples has not changed.
They are packed into \gls{udp} datagrams, sent across the network, and reassembled by the host \gls{cpu}, exactly as they were a decade ago.

Ethernet is the right medium for this.
It is commodity hardware, it is routable, and it scales from a laboratory bench to a distributed measurement facility and from meters to kilometers.
A \gls{usb} or \gls{pcie}-attached \gls{sdr} serves a single host, cannot be operated remotely, and -- in the case of \gls{pcie} -- typically incurs further restrictions regarding compatible software and hardware.
Similarly, specialized high-speed serial protocols like Aurora or Interlaken depend on rare and thus costly host interfaces.
The question is not whether to carry samples over Ethernet, but how.

The current answer fails in two independent ways.
The host must process every packet, and at 8\,GB/s it has roughly one microsecond per packet, which is what the socket path itself consumes on a modern operating system.
The host must also reassemble the packets into a contiguous sample stream, which makes every sample byte cross the memory bus five times.
Two X440 channels then generate 80\,GB/s of memory traffic before the application performs any processing at all, which is almost the entire bandwidth of a contemporary desktop host.

Neither is a tuning problem.
The transport requires the application to own its buffers, its packet handling, and its flow control, which necessitates expertise in operating systems rather than in radio.
The literature shows what follows: capable groups fall back to periodic burst capture rather than clear the requirement, and they do so at 200\,MSa/s.
Ten times that rate is now available in a shipping product and more is already on the horizon.
The \gls{uhd} has responded by offering an interface that hands the transport back to the user outright.

We propose to stop building the transport and to use one that already exists.
\Gls{rdma} lets a network interface move data directly into and out of registered host memory, performing packetization, reassembly, and \gls{dma} in hardware.
\Gls{rocev2} provides it over ordinary Ethernet, natively on 100+~\gls{gbe} class network interfaces and in software on Linux for lower transfer rates.
The \gls{sdr} writes received samples into host memory and reads samples for transmission out of it.
The host \gls{cpu} is not in the data path in either direction, leaving it for its intended purpose: processing samples.

This article makes the following contributions.
We quantify both limits of the present transport across four generations of converter front end, and show that kernel bypass addresses one of them and not the other.
We describe an \gls{sdr} streaming architecture built on \gls{rocev2}, in which the samples reside in registered host memory, the connection parameters are configured out of band, and no \gls{rdma} connection manager is required on the device.
We invert the direction of transmit streaming: rather than the host intricately pushing samples into a buffer that must never under- or overflow, the \gls{sdr} reads them exactly when it needs them.
We show that the same mechanism aggregates multiple links into one contiguous sample stream, which is what rates exceeding a single Ethernet link require.
We argue that the resulting \gls{sdr}-side implementation is tractable, adding 32 bytes of transport header to packets that already carry Internet Protocol and \gls{udp} headers today.

\autoref{sec:today} examines the present transport and its limits.
\autoref{sec:rdma} introduces \gls{rdma} and \gls{rocev2} to the extent that the architecture requires.
\autoref{sec:architecture} presents the architecture itself.
\autoref{sec:outlook} discusses what the transport enables.
This article presents the architecture and its rationale; comparative measurements are the subject of ongoing work.

\section{How SDRs Stream Samples Today}
\label{sec:today}

\subsection{The Prevailing Model}
\label{sec:prev-model}

\Gls{sdr} sample transport carries a continuous stream of \gls{iq} samples between the data converters and the host.
Ethernet requires that this stream be segmented into packets at the sender and reassembled at the receiver.
We refer to this as the packetize--reassemble model.
Today, every Ethernet-attached \gls{sdr} operates that way.

Several standards govern what those packets contain.
\Gls{vita49} and its \gls{difi} profile define sample encoding, timestamps, context metadata, and stream identification.
All of them specify payload framing.
\Gls{chdr} serves the same purpose in \gls{uhd}, as does \gls{ecpri} in fronthaul applications.
All of them specify payload framing.
None of them specifies how the samples reach the memory of the application that consumes them.
That step is left to the host operating system and to the application itself.

\subsection{The Tuning Tax}
\label{sec:tuning-tax}

Continuous loss-free streaming at gigasamples per second requires the host to be configured for it.
Interrupt affinity and thread placement must be pinned to the \gls{cpu} cores serving the \gls{nic}.
\Gls{nic} ring descriptor counts and socket buffer sizes should be raised well above their defaults.
Cores carrying the receive path should be isolated from the scheduler, and a real-time kernel may be necessary.
None of these requirements concerns radio.
Omitting any one of them increases the likelihood of sample loss.

The consequences are visible in the literature.
Where pre-processing sample data on the \gls{sdr} is not feasible algorithmically, due to \gls{fpga} resource constraints, or for lack of \gls{fpga} or systems programming knowledge, researchers have to accept the available performance threshold.
Rather than streaming continuously, they resort to periodic burst sampling to reduce the demands on the host~\cite[pp.~166\,f.]{nuss21_dissertation},~\cite{diouf21_tim_x310_burst_200ma, merlo24_tmtt_x310_burst_200msa}.

Burst sampling lowers the sustained data rate, but it does not change the cost of moving a single packet.
Even a relatively short sub-millisecond burst results in hundreds of packets.
The fallback to burst operation is therefore a concession to host performance rather than a solution, and it is unavailable to applications that require zero measurement dead time.

\subsection{Who Owns the Buffer?}
\label{sec:buffer-ownership}

\Gls{uhd} requires the application to pass a pointer into the streaming \gls{api}.
The buffer behind that pointer is allocated by the user application, and its properties are therefore the user's responsibility.
It must be backed by hugepages, locked into physical memory, allocated on the \gls{numa} node serving the network interface, and sized against the bandwidth--delay product of the link.

The \gls{api} enforces none of this.
A buffer that violates every one of these requirements is accepted and works.
It continues to work until the sustained rate rises or the scheduler makes an unfavorable decision, at which point samples are lost.
The failure surfaces long after the cause, and the diagnostic path leads through host configuration rather than through the radio.

The information needed to make these decisions correctly is available to the driver.
It knows the sample rate, the link rate, and the topology of the host.
The application, in the general case, does not.
The abstraction boundary is drawn such that the party with less information is required to make the decision.

\subsection{The Per-Packet Budget}
\label{sec:packet-budget}

The tuning requirements of \autoref{sec:tuning-tax} exist because the host must process every packet individually.
Dividing the sample rate by the packet payload size yields the packet rate, and its reciprocal is the time available to the host per packet.
\autoref{tab:budget} lists both for a range of converter front ends at a payload of 8192\,B.\footnote{%
Jumbo frames permit a link \gls{mtu} of 9000\,B, but \gls{uhd} caps its own \gls{mtu} at 8000\,B, which yields 1996 samples per packet on an X310.}

Processing a single packet through the socket \gls{api} costs the host on the order of one to two microseconds.
The X310 and X410 therefore leave the host an order of magnitude more time than it needs.
The X440 does not.
At 1.02\,\textmu s per packet it lands inside the range that the socket path itself consumes, leaving nothing for the application.
The converters beyond it miss by an order of magnitude or more.

The Versal RF front end additionally produces 64\,GB/s from a single channel, which is five times what a 100~\gls{gbe} link carries.
No host-side optimization addresses that.

\begin{table}
\caption{Per-packet time budget at 8192\,B payload and 4\,B per \gls{iq} sample.}
\label{tab:budget}
\centering
\begin{tabular}{lrrrr}
\toprule
SDR front end & Rate/ch. & Throughput & Packets/s & Budget \\
\midrule
USRP X310    & 200\,MSa/s  & 0.8\,GB/s & 98\,k   & 10.2\,\textmu s \\
USRP X410    & 500\,MSa/s  & 2\,GB/s   & 244\,k  & 4.10\,\textmu s \\
USRP X440    & 2\,GSa/s    & 8\,GB/s   & 977\,k  & 1.02\,\textmu s \\
RFSoC ZU47DR & 2.5\,GSa/s  & 10\,GB/s  & 1.22\,M & 819\,ns \\
Versal RF    & 16\,GSa/s   & 64\,GB/s  & 7.81\,M & 128\,ns \\
\bottomrule
\end{tabular}
\end{table}

\subsection{Copy Amplification}
\label{sec:copy-amplification}

The per-packet budget is not the only limit.
Reassembling packets into a contiguous sample stream costs memory bandwidth, and that cost scales with the sample rate rather than with the packet rate.

\begin{table}[!t]
\caption{Memory bus crossings per sample byte. For the socket and bypass paths the \gls{dma} write lands in a driver-owned buffer; for \gls{rdma} it lands in the registered buffer of the application.}
\label{tab:copies}
\centering
\begin{tabular}{llccc}
\toprule
Stage & Operation & Socket & Bypass & \gls{rdma} \\
\midrule
\Gls{nic} \gls{dma}      & write      & 1 & 1 & 1 \\
Kernel to user space     & read/write & 2 & 0 & 0 \\
Reassembly in user space & read/write & 2 & 2 & 0 \\
\midrule
\multicolumn{2}{l}{Total} & 5 & 3 & 1 \\
\bottomrule
\end{tabular}
\end{table}

Consider the path a sample byte takes from the network interface to the application.
The interface writes the packet into a kernel receive buffer by \gls{dma}, which is one write.
The socket \gls{api} copies the payload into the buffer supplied by the application, preserving packet boundaries, which is one read and one write.
The application then separates the stream headers from the samples and assembles them into a contiguous block, which is another read and write.
Each sample byte therefore crosses the memory bus five times, as summarized in \autoref{tab:copies}.\footnote{%
The second copy can in principle be avoided with scatter-gather reception through \texttt{recvmmsg()}, which \gls{uhd} does not implement.
The first copy remains regardless.}

\autoref{tab:budget} lists 8\,GB/s for a single X440 channel.
Two channels amount to 16\,GB/s of samples and 80\,GB/s of memory traffic before the application performs any processing.
A contemporary desktop host with dual-channel DDR5-6400 provides roughly 100\,GB/s in total.
A single Versal RF channel at 64\,GB/s reaches 320\,GB/s and exceeds the roughly 307\,GB/s of an eight-channel server.
The DRAM figures are theoretical peaks.
The traffic figures are worst case and assume that every crossing reaches \gls{dram}.\footnote{%
    Where the reassembly buffer is small enough to remain resident in the last-level cache, a substantial part of the traffic never reaches \gls{dram}, and configurations that fail with a larger buffer become feasible.
    Further platform-specific optimizations, such as direct cache injection, may reduce it further, but lie beyond the scope of this manuscript.%
}

\subsection{Kernel Bypass Is Not Enough}
\label{sec:kernel-bypass}

Kernel bypass frameworks such as \gls{dpdk} and \gls{afxdp} are the established answer to the per-packet cost.
They map the receive queues of the network interface into user space and let the application poll them directly, which removes the per-packet system call and the interrupt that accompanies it.
The per-packet budget of \autoref{sec:packet-budget} ceases to be the binding constraint.

The memory cost is only partly addressed.
Bypass eliminates the copy from kernel to user space, but the application still receives packets rather than a sample stream, and it must still assemble them into a contiguous block.
Three crossings per sample byte remain of the five in \autoref{sec:copy-amplification}.
Reducing memory traffic by two fifths does not change the conclusion for the front ends at the lower end of \autoref{tab:budget}.

Bypass also moves in the wrong direction with respect to \autoref{sec:buffer-ownership}.
The application now owns the memory pools, the queue configuration, and the polling threads in addition to the sample buffer.
\Gls{dpdk} requires a poll-mode driver bound to the interface, which removes it from the network stack of the operating system and from the tooling built around it.
The expertise required of the user increases rather than decreases.

Neither framework addresses transmission.
Both provide a mechanism to send packets, but the decision of when to send remains with the application, which must anticipate the consumption of the transmit chain without visibility into it.
Neither addresses aggregation across multiple links.

\subsection{Last Resort: Bypass the Abstraction}
\label{sec:last-resort}

\Gls{uhd} offers a feature it calls remote streaming, in which an Ethernet-attached device sends samples to a destination other than the controlling \gls{uhd} session~\cite{uhd-manual-stream}.
The \gls{rx} streamer remains in place as a proxy for stream commands, but the samples arrive at a socket the receiving application opens and manages itself.

What that user application inherits is instructive.
Flow control is disabled by default, and the device streams at its configured rate whether or not the destination keeps up.
Enabling it requires switching to full-packet mode and implementing the \gls{rfnoc} flow control response protocol at the destination.
Full-packet mode in turn requires the destination to parse and strip the \gls{chdr} header.
Automatic \gls{mtu} detection is unavailable, and an incorrect value causes fragmentation and data loss.
An overrun is reported to the controlling \gls{uhd} session, not to the destination receiving the samples.

The interface is a reasonable engineering decision.
Within the packetize--reassemble model there is no other way to give an application the control it needs at these rates.
It exists because the model has reached its limit, and the remaining option was to return the transport to the user.

Remote streaming is available for reception only.
Implementing the equivalent for transmission is possible in principle, but it would hand the user application an even harder problem.
Samples must reach the converter fast enough to keep its buffer from running dry and slowly enough to keep it from overflowing.
Reliably satisfying this hard real-time requirement increases the implementation burden substantially, necessitating flow control messages, precise sub-millisecond scheduling logic, and pacing of sent sample packets.

\section{RDMA for Sample Streaming}
\label{sec:rdma}

\subsection{Precedent, and How It Differs}
\label{sec:precedent}

Data rates comparable to those of \autoref{tab:budget} are not unique to \gls{sdr}.
Radio astronomy correlator backends and the \gls{felix} readout system at CERN move data at similar rates in real time, and both employ \gls{rdma} to do so~\cite{radioastronomy-rdma,felix}.

Such systems typically reduce their data before it reaches the network.
Pipelined processing in the front end selects, integrates, or filters the raw stream, and only the reduced product is transmitted.
The decision as to what is relevant is made ahead of the transfer.

An \gls{sdr} front end cannot make that decision in the general case.
What is relevant is defined by the application on the host, which is precisely what makes the radio software-defined.
The transport must therefore be able to carry the full sample stream, whether or not a given application permits reduction ahead of it.
The authors are not aware of an application of \gls{rdma} to \gls{sdr} sample streaming in the public record.

\subsection{What RDMA Changes}
\label{sec:rdma-changes}

\Gls{rdma} moves the transfer into the network interface.
The interface reads the data out of host memory by \gls{dma} and packetizes it in hardware, or it reassembles incoming packets in hardware and writes the result into host memory by \gls{dma}.
The \gls{cpu} takes no part in moving the bytes.
There is no system call per packet, no interrupt per packet, and no reassembly loop, so the per-packet budget of \autoref{sec:packet-budget} no longer applies.
The samples are written to the location where the application reads them, which is the \gls{rdma} column of \autoref{tab:copies}.

Before streaming begins, the memory that will hold the samples is registered with the network interface.
Registration locks the pages into physical memory and makes the region reachable from the network.
It does not choose the allocation, which remains the responsibility of whoever performs it, so the properties of \autoref{sec:buffer-ownership} still have to be right.
What changes is that the samples arrive in the registered region and are read from it directly.
The buffer can therefore be allocated by the driver and handed to the application, instead of being allocated by the application and handed to the driver.
The host then configures the \gls{sdr} with the information it needs to reach the region.
This happens once.

Once the host triggers a stream, the \gls{sdr} carries it out on its own initiative.
To deliver received samples, the \gls{sdr} writes them into the registered region of the host.
To obtain samples for transmission, the \gls{sdr} reads them out of the registered region.
The network interface of the host answers such a read from the registered memory in hardware.
No thread runs, no call is made, and the host \gls{cpu} remains unaware that the transfer took place.

The host therefore leaves the data path in both directions.
It defines where the samples live and when streaming starts, and the \gls{sdr} determines when each transfer occurs.

Packetization and reassembly do not disappear.
They move into the network interface, which performs them at line rate in dedicated logic.

\subsection{RoCEv2 in Practice}
\label{sec:rocev2}

\Gls{rocev2} carries \gls{rdma} over ordinary Ethernet and \gls{udp}~\cite{ibta-a17}.
Its payload is limited to 4096 bytes, so the packet lengths in current use are split across two packets.
Counting the additional transport headers and the doubled framing, the approximate total wire overhead for 8192 bytes of samples rises from 1\,\
Although the packet rate doubles, all \gls{rocev2} packets are processed in hardware and never reach the network stack, having no effect on \gls{cpu} load.

Because the packets are ordinary routable datagrams, \gls{rocev2} traffic shares the interface, the subnet, and the switching fabric with everything else.
The interface remains available to the operating system and to any other application using it.
\Gls{dpdk} offers no such coexistence: its poll-mode drivers bind the interface and remove it from the network stack, unless the hardware supports bifurcation.

\Gls{rocev2} offload is a standard feature of 100~\gls{gbe} class network interfaces.
Where such an interface is not available, Linux provides \texttt{rxe}, a software implementation that operates over any Ethernet device and presents the same interface to the application~\cite{rxe-man}.
It reintroduces host \gls{cpu} cost and may not reach gigasample rates.
It is nonetheless not the same as implementing a transport in the application, because the work is done by kernel code that is maintained and optimized by experts and runs isolated from the application rather than competing with it for the same threads.
At 10, 25, and 40~\gls{gbe} this is likely sufficient, obviating a fallback, non-\gls{rdma} sample path in any \gls{sdr} driver.

\section{An RDMA-Native Streaming Architecture}
\label{sec:architecture}

\subsection{Setup and the Control Plane}
\label{sec:setup}

Before streaming begins, the host reserves the memory that will hold the samples and registers it with its \gls{nic}.\footnote{The properties of \autoref{sec:buffer-ownership} apply unchanged, but the allocation is now performed by the driver, which knows the sample rate, the link, and the topology of the host.}
Several separate registrations may be established.
The host then selects a registration, or a subset of one, and configures the \gls{sdr} accordingly.
It thereby instructs the \gls{sdr} where to write the samples it receives, or where to read the samples it is to transmit.

The \gls{sdr} streams accordingly, its network logic computing the address for every sample packet it sends or requests.
We use network logic to denote the packet-generating logic of the \gls{sdr}, implemented in an \gls{fpga} or an \gls{asic}, as distinct from the \gls{nic} of the host.
The configuration determines how the memory is used: once, for a burst of limited duration, or cyclically as a circular buffer for continuous streaming.
By queuing multiple streaming commands, the \gls{sdr} can also realize continuous streaming through consecutive bursts, using arbitrary memory regions chosen opportunistically by the host.

The configuration itself occurs out of band.
It happens occasionally, it carries no samples, and it takes no part in the transfers that follow.
A register-level protocol tunneled over Ethernet, such as \gls{axioe}~\cite{kamp17_axioe}, enables a lean implementation optimized for resource-constrained \glspl{asic} and \glspl{fpga}.
The host writes the relevant protocol parameters directly into the network logic, which obviates the state machines that a full \gls{rdma} connection manager would require.
The network logic reduces to generating static headers, incrementing a subset of their fields, such as addresses and sequence numbers, and computing checksums.

The host defines when a stream starts and when it ends, by wall-clock time or in response to an event, and configures the network logic accordingly.
Everything after that is carried out by the network logic on its own.

\subsection{Receive: The SDR Sends}
\label{sec:receive}

The network logic segments the sample stream, generates the packet headers, and sends the packets.
The \gls{nic} of the host reassembles them and writes the payload into the registered memory region by \gls{dma}.

What arrives in that region is the sample stream itself.
The payload of consecutive packets is stored sequentially and without gaps, with no headers preceding, separating, or following it.
The application reads a contiguous block of samples, and no copy takes place within host memory.
This is the \gls{rdma} column of \autoref{tab:copies}.

The host is passive throughout.
No thread runs, no call is made, and no interrupt is raised per packet.
Where the application wants to be notified of progress, the network logic can optionally pass immediate data into the \gls{rdma} write, which the host is informed of and must actively consume.
This can occur for every packet or, for performance reasons, for a subset of them, at the discretion of the host.

Metadata is not interleaved with the samples, as that would reintroduce the separation \autoref{sec:copy-amplification} charges for.
Timestamps and stream state travel on a separate path, either pulled by the host over the configuration channel or pushed by the \gls{sdr} over a conventional socket or \gls{afxdp}.
Their rate is low enough that the cost of either is irrelevant.

\subsection{Transmit: The SDR Reads}
\label{sec:transmit}

\autoref{sec:last-resort} ended on a hard real-time problem: the host must deliver samples fast enough to keep the converter buffer from running dry and slowly enough to keep it from overflowing, without being able to observe it directly.
The architecture removes that problem by reversing the direction of the transfer.

The network logic requests the samples it needs, when it needs them.
It sends request packets addressed to the registered memory region of the host.
The \gls{nic} of the host processes each request, reads the requested samples from that region by \gls{dma}, packetizes them, and sends them back.
The network logic reassembles the arriving payload into its converter buffer.
The host \gls{cpu} takes no part in this, exactly as it takes no part in reception.

The party that knows the fill level of the converter buffer is now the party that decides when to transfer.
The network logic sets the rate and the rhythm of its own requests and adapts them continuously to that fill level.
No flow control protocol is required in either direction, because nothing needs to be told anything.

The lead time of a request follows from the round-trip time between issuing it and the requested samples arriving.
The network logic measures that time, once, repeatedly, or continuously, and shifts subsequent requests accordingly.
The lead time thereby becomes a measured quantity rather than a configured guess, and it tracks changes in the network without intervention.

What \autoref{sec:last-resort} required of the user application was flow control messages, precise sub-millisecond scheduling logic, and pacing of transmitted packets.
None of the three remains.

\subsection{Tracking the SDR, and Handling Loss}
\label{sec:tracking}

The host is not in the data path, but it still needs to know where in the registered region the \gls{sdr} is currently writing or reading.
For real-time applications this must be known precisely, since the margin the host keeps between itself and the \gls{sdr} is latency it adds to the system.
A clock synchronized between host and \gls{sdr} provides this: from the synchronized time and the configured sample rate, the host derives the position the \gls{sdr} has reached.

The region can also be observed directly.
Initializing it at defined addresses with distinguishable placeholder values lets the host detect progress by their disappearance, since arriving samples overwrite them.
A placeholder still present where samples were expected localizes a gap exactly, without inspecting a single packet header and without any interrupt load.

Such gaps will occur.
Bit errors arise eventually even on a point-to-point link with forward error correction, and a corrupted packet is discarded.
Loss must therefore be handled, and on an \gls{fpga} or \gls{asic} it must be handled within a few megabytes of fast on-chip memory.

Reliable \gls{rdma} transports retransmit lost packets and unreliable ones do not.
Under a reliable transport the \gls{nic} of the host reports the loss, and the network logic resends the affected samples if it still holds them, or placeholder values, typically zeros, if it does not.
Under an unreliable transport nothing is resent, and the host fills the gap with zeros itself, which the placeholder mechanism has already located for it.
On transmit the same choice applies: the network logic re-requests the missing samples, or substitutes a virtual packet of zeros of equal length.

Substituting zeros is what keeps the device-side implementation small.
Retransmitting from the source requires holding every sent packet until it is acknowledged, and at these rates that buffer is the bandwidth--delay product of the link, which may easily exceed the on-chip memory of the devices in question.
Substitution also keeps the sample stream aligned in time, which retransmission under a hard deadline does not guarantee.

\subsection{Scaling Beyond a Single Link}
\label{sec:aggregation}

The data rate of a single channel can exceed the rate of a single network link, as \autoref{tab:budget} shows for the highest converter rates.
No host-side measure addresses this, and the devices concerned provide several links for that reason.

The network logic distributes the packets of one sample stream across those links.
Each packet carries the address at which its payload belongs, so the samples are placed correctly regardless of which link they arrived on and in which order.
Reassembly is unaffected because placement follows from the address rather than from arrival.
Bandwidth scales linearly with the number of links, and no copy is reintroduced.

Transmission works the same way.
The network logic distributes its requests across the links, and each \gls{nic} of the host answers over the link on which the request arrived.

The mapping need not be one to one.
A single \gls{nic} may serve several links of the \gls{sdr}, or one link may be split across several \glspl{nic}, which is useful when the two differ in rate.
The registered regions may likewise be distributed across several \glspl{nic}, several storage devices, or several hosts, so that one channel is spread over multiple receivers for load distribution.

Several independent sample streams are distinguished by their packet headers and placed in separate regions.
A multi-channel \gls{sdr} thereby delivers each channel to the memory the host assigned to it, over any combination of links.

\subsection{What the Application Is Left With}
\label{sec:application}

The requirements of \autoref{sec:tuning-tax} and \autoref{sec:buffer-ownership} have not disappeared, but the application no longer needs to meet them.
The driver reserves and registers the memory and hands it to the application, rather than receiving a pointer whose properties it cannot check.
Hugepage backing, page locking, and \gls{numa} placement are settled once, by the party that knows the sample rate, the link, and the topology of the host.

The application is left with a region of memory that fills with samples, or that it fills with samples.
It does not see packets, so it does not size rings or tune interrupts.
It does not pace anything, because the \gls{sdr} asks for what it needs.
It does not implement flow control, because there is none to implement.

What remains of the transport is performed by the \gls{nic}, by kernel code, or by the network logic of the \gls{sdr}.
All three are maintained and optimized by specialists, and the Linux \gls{rdma} stack in particular receives continuous attention.
Implemented correctly, the architecture places multi-gigasample per second streaming within reach of a few lines of Python.
\Gls{rdma} absorbs the packet handling, the driver absorbs the implementation, and what is left for the author of the application is the signal processing itself.

\section{Outlook}
\label{sec:outlook}

\Gls{rocev2} is the natural transport for this architecture, but it is not the only one that realizes it.
What the architecture requires is that packetized samples be written into, or read out of, a pre-registered memory region by \gls{dma}, that the payload of two or more consecutive packets come to rest sequentially and without headers, and that no copy occur within host memory.
\Gls{iwarp} and InfiniBand satisfy these requirements and are direct substitutions.

A variant of the architecture dispenses with addresses altogether.
\autoref{sec:aggregation} placed samples correctly because every packet carried the address at which its payload belonged.
The same placement follows from the chronological order of the packets, or from sequence numbers in their headers, and no address is needed to obtain it.

This requires no special hardware.
Any \gls{nic} capable of scatter-gather reception can separate the packet header from the payload and deposit the two in different buffers, a capability marketed variously as header-data split or \gls{shampo}.
Posting receive buffers whose payload segments are adjacent in memory, at a fixed payload size, causes the payloads of consecutive packets to come to rest sequentially and without headers between them.
The result in host memory is identical to that of \autoref{sec:receive}, and it is reached without a single copy.
The packets themselves are then ordinary InfiniBand SEND operations, carrying immediate data where the host is to be notified.
Loss shifts rather than perforates the stream: a missing packet leaves no gap, because the payload that follows occupies the slot intended for it.
The shift is detectable from the immediate data or from the split-off header, and it is repaired by moving the affected payload once.
At realistic loss rates the cost of doing so is negligible against the transfer it corrects.

A configurable network interface reaches the same result differently, deriving the write position from the sequence rather than from the buffers posted ahead of it, and permitting that configuration to be adjusted while streaming continues.
Either way the packet headers shorten and the network logic reduces further, since addresses no longer have to be computed.

The registered region need not reside in host memory at all.
It may be located in the memory of a compute accelerator, so that samples arrive where they will be processed, or on a non-volatile storage device, so that a recording is written without passing through main memory.
The transfer then takes place directly between the \gls{nic} and the peripheral over \gls{pcie} peer-to-peer, and the bandwidth accounting of \autoref{sec:copy-amplification} no longer involves the host at all.

Where an application does permit its samples to be reduced before they reach the network, the network logic is a natural place to do it.
Filtering, decimation, integration, buffering, and selective transmission implemented alongside the packet generation reduce what has to be transferred.
This does not change the architecture; it changes only how much of the stream it carries.
The transport remains capable of carrying all of it, which \autoref{sec:precedent} identified as the property that distinguishes \gls{sdr} from the systems that established this pattern.

\section{Conclusion}
\label{sec:conclusion}

Sample rates have outgrown the transport that carries them.
The host cannot process a million packets per second within the time each packet allows, and it cannot reassemble tens of gigabytes per second into a contiguous stream without spending the memory bandwidth the application needs for its own work.
Neither limit is a matter of configuration, and kernel bypass addresses only the first.

The transport this article describes delegates the problem to hardware that already solves it.
The \gls{sdr} writes the samples it receives into memory the host registered in advance, and it reads the samples it transmits out of that same memory when its converter needs them.
The host defines where the samples live and when streaming begins, and takes no further part.
Each sample byte crosses the memory bus once.

The transmit path is the more consequential of the two.
Pushing samples toward a converter buffer that cannot be observed is a hard real-time problem that has been imposed on every user of a high-rate \gls{sdr}.
Letting the \gls{sdr} ask for what it needs removes it, and with it the flow control, the scheduling logic, and the pacing that the problem demanded.
The implementation burden falls accordingly, and what is left of gigasample \gls{sdr} development is the radio work that really matters rather than systems programming and memory management boilerplate.

What remains to be shown is measurement.
The architecture is implemented as a proof of concept, and a comparison against the transports of \autoref{sec:today} across the front ends of \autoref{tab:budget} is the subject of ongoing work.

\section*{Acknowledgment}
The authors used artificial intelligence (Anthropic Claude Opus~5) to draft the prose of this manuscript.
The document structure, research question, the technical substance, the architecture, and all judgments presented are the authors' own.
The authors reviewed the complete text, verified its technical content, and take full responsibility for it.

\IEEEtriggeratref{3}
\bibliographystyle{IEEEtran}
\bibliography{paper}

\end{document}